\documentclass[preprint,12pt,authoryear]{elsarticle}
\usepackage{amssymb}
\usepackage{lineno}
\usepackage{booktabs}
\usepackage{bm}
\usepackage{multirow}
\usepackage[table,xcdraw]{xcolor}
\usepackage{xr-hyper}
\usepackage{hyperref}
\usepackage{gensymb}
\usepackage{xcolor}
\usepackage{soul}
\usepackage[normalem]{ulem}
\usepackage{booktabs}
\usepackage{multirow}

\newcommand{\mat}[1]{\bm{\mathrm{\MakeUppercase{#1}}}}
\renewcommand{\vec}[1]{\bm{\mathrm{#1}}}
\newcommand{\real}[1]{\mathbb{R}^{#1}}

\journal{Energy}

\begin{document}
\begin{frontmatter}

\title{Increasing the skill of short-term wind speed ensemble forecasts combining forecasts and observations via a new dynamic calibration}

\author[DICCA]{Gabriele Casciaro} \ead{gabriele.casciaro@edu.unige.it} 

\author[DICCA,INFN]{Francesco Ferrari} \ead{francesco.ferrari@edu.unige.it}

\author[MALGA,DICCA]{Daniele Lagomarsino Oneto} \ead{daniele.lagomarsino@edu.unige.it}

\author[DICCA]{Andrea Lira-Loarca} \ead{andrea.lira.loarca@unige.it}

\author[DICCA,INFN]{Andrea Mazzino\corref{cor}} \ead{andrea.mazzino@unige.it}

\affiliation[DICCA]{organization={DICCA, Department of Civil, Chemical and Environmental Engineering. University of Genoa},
            addressline={Via Montallegro 1}, 
            city={Genoa},
            postcode={16145}, 
            state={Genoa},
            country={Italy}}
\affiliation[INFN]{
            organization={INFN, National Institute of Nuclear Physics, Genoa section},            addressline={Via Dodecaneso 33}, 
            city={Genoa},
            postcode={16146}, 
            state={Genoa},
            country={Italy}}
\affiliation[MALGA]{organization={MALGA, Machine Learning Genoa Center. University of Genoa},
            addressline={Via Dodecaneso 35}, 
            city={Genoa},
            postcode={16146}, 
            state={Genoa},
            country={Italy}}

\cortext[cor]{Corresponding author}

\begin{abstract}
All numerical weather prediction models used for the wind industry need to produce their forecasts starting from the main synoptic hours 00, 06, 12, and 18 UTC, once the analysis becomes available. The six-hour latency time between two consecutive model runs calls for strategies to fill the gap by providing new accurate predictions having, at least, hourly frequency. This is done to accommodate the request of frequent, accurate and fresh information from traders and system regulators to continuously adapt their work strategies. Here, we propose a strategy where quasi-real time observed wind speed and weather model predictions are combined by means of a novel Ensemble Model Output Statistics (EMOS) strategy. The success of our strategy is measured by comparisons against observed wind speed from SYNOP stations over Italy in the years 2018 and 2019. 
\end{abstract}




\begin{highlights}
\item   First attempt to couple in an efficient and economic way real-time data and ensemble predictions.
\item   First assessment of the added value of real-time observations in a wind calibration.
\item   Real-time data can be easily and economically ingested in an EMOS-based calibration.
\item   Ingestion of real-time data produces noticeable benefits vs. static calibrations.
\item   Real-time data provide added value to the whole  wind predictive probability density.
\end{highlights}


\begin{keyword}

wind forecasting \sep probabilistic forecasting \sep dynamic forecast calibration \sep ensemble model output statistics \sep wind forecast based on real-time conditions \sep Numerical Weather Prediction models

\end{keyword}

\end{frontmatter}


\section{Introduction}
\label{Introduction}
Global cumulative installations of onshore and offshore wind are expected to exceed 1 TW before 2025 \citep{council2021gwec}. This means that the contribution of wind power in power systems is becoming increasingly important. The downside is that detailed schedule plans and reserve capacity must be properly set by power system regulators \citep{impram2020challenges} facing the intrinsic problem of the highly intermittent nature of wind, making this very hard to predict. The accuracy of wind forecasts thus becomes an issue of paramount importance for the wind industry.\\
In a recent work by \cite{casciaro2021comparing}, a novel accurate Ensemble Model Output Statistics (EMOS) strategy for calibrating wind speed/power forecasts from an Ensemble Prediction System (EPS) has been proposed and its superiority when compared against more parsimonious strategies in the 0-48 h look-ahead forecast horizon clearly emerged. However, because all global weather models start their run from analysis corresponding to the main synoptic hours 00, 06, 12, and 18 UTC, weather predictions (of any forecast horizons) necessarily remain frozen for six hours. This limitation is in sharp contrast with the needs of power system regulators, as well as of traders for marketing wind energy, who need to adapt their strategies hour after hour in a quasi-continuous way. It is thus very important to propose accurate strategies which give fresh information on the wind speed in a given location continuously evolving between two consecutive main synoptic hours. Proposing a strategy with such characteristics is the main aim of the present paper. In plain words, we propose a novel, parsimonious,  dynamic EMOS strategy where the parameters entering the EMOS predictive probability density function now also depends on real-time (or quasi-real time) observed wind speed data.
Our strategy combines two well-known advantages for  wind prediction by
physical methods (i.e.\ based on numerical weather prediction, NWP, models) and statistical methods (i.e.\ statistical models based on time-series of past measured observables and/or observation-driven machine-learning (ML) based methods). 
While the former methods are suitable for predicting wind (and, more generally, any meteorological observable) in the  forecast horizon larger than, say, 6 hours, the latter strategies turn out to have greater skills for short-term forecasting, say, less than a few hours ahead. For a review of those methods and discussions of their skills, see, e.g., \cite{soman2010review} and the references therein. We will show that combining quasi real-time  observed data and predictions via a suitable EMOS strategy provides an optimal assembly strategy which outperforms the single aforementioned strategies when acting separately. This turns out to be the case both in terms of point indices (the standard NMAE and correlation coefficient) and
via suitable statistical indices to assess the whole probability density function of the (calibrated) ensemble forecast error.\\
For our dynamic strategy to be used operatively, the observed wind speed on the site of interest must be available in real/quasi-real time, other than as a record of past observations for implementing the static calibration.

The paper is organized as follows. In Sec.\ \ref{Sec:obs} we introduce the wind speed observations dataset used both to train our calibration algorithm and for its testing. Sec.\ \ref{sec:EPS} presents the forecast data from the Ensemble Prediction System (EPS) of the ECMWF.   Sec.\ \ref{sec:EMOS0} provides a quick review of the standard Ensemble Model Output Statistics (EMOS). A very recent generalization of standard EMOS to account for nonlinearities is described in Sec.\ \ref{sec:EMOS+}. The algorithm we propose to ingest real-time observed data in a EMOS strategy is described in Sec.\ \ref{sec:algo}.  The new algorithm is tested in different respects in Secs.\ \ref{Sec:eval-step} and \ref{Sec:assess}.
The final section is devoted to draw some conclusions and perspectives.

\section{Wind data}
\label{Wind data}
\subsection{Observed data from SYNOP stations}
\label{Sec:obs}
\begin{figure}[h!]
\includegraphics[width=0.8\textwidth]{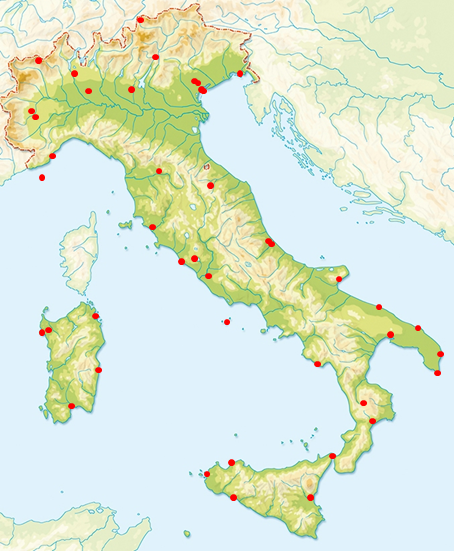}
\centering
\caption{Location of a selected group of Italian SYNOP stations used in the present study for forecast calibration and testing. Colors are coded according to the orography elevation.}
\label{fig:stazioni_synop_italia}
\end{figure}

From 2018 to 2019, SYNOP meteorological data were collected at 42 locations across Italy.
According to ICAO specifications, the SYNOP anemometers record wind speed (knots) as an average over 10 minutes at a nominal measurement height of 10 meters a.g.l. \citep{icao2007meteorological} with hourly frequency.
The stations are arranged fairly uniformly across the peninsula, as shown in Figure\ \ref{fig:stazioni_synop_italia}. Data have been split using 2018 as training set and 2019 as test set.

\subsection{Forecast data from the ECMWF Ensemble Prediction System (EPS)}
\label{sec:EPS}

The Ensemble Prediction System (EPS) from the European Centre for medium-range weather forecasts (ECMWF), is made up of 51 members: a control forecast with no perturbations and 50 forecasts created by adding small perturbations to the best-known initial condition, using a mathematical formulation based on singular vector decomposition and stochastic parameterization to represent model uncertainties \citep{buizza1995optimal, leutbecher2008ensemble}.\\
The EPS used in this study has spectral triangular truncation with a cubic-octahedral grid Tco639 and 91 layers with a top of atmosphere pressure of 0.01 hPa \citep{buizza2018ensemble} and a resolution of about 18 km \citep{persson2001user}.\\
Although four runs per day are made available from the ECMWF, in the present study the sole EPS runs starting from the 00 UTC analysis have been considered
in  the first 48-hour look-ahead forecast horizon.\\
Although there are robust evidences \citep{molteni1996ecmwf, montani2019performance} that observed  mean and variance correlate well with the EPS ensemble mean and variance, all EPS members frequently underestimate and are underdispersive with respect to observations. Then EPS forecasts must be properly calibrated, especially for close-to-surface meteorological data. The surface wind speed is just a relevant example considered in the present paper.


\section{The static calibration strategy}
\label{sec:static}
We provide in this section a brief review of the so-called Ensemble Model Output Statistics (EMOS) through which a raw ensemble forecast can be accurately transformed into a predictive probability density function, with a simultaneous  correction for biases (additive and multiplicative) and dispersion errors \citep{gneiting2005calibrated,thorarinsdottir2010probabilistic}.
A generalization recently proposed by \cite{casciaro2021comparing} accounting for nonlinear relationships between predictands and both predictors, and other weather observables used as conditioning variables  will be also reviewed. \\
Both strategies share the common feature of being static, meaning that they do not exploit real-time (or quasi-real time) observed data. 


\subsection{The standard Ensemble Model Output Statistics}
\label{sec:EMOS0}
The standard EMOS strategy, here denoted as EMOS$_0$, is a generalization of multiple linear regressions, also known as Model Output Statistics (MOS) \citep{glahn1972use}, which is commonly used to calibrate deterministic forecasts.\\
Let us consider $M$ ensemble member forecasts, $X_1, \cdots, X_M$, of a univariate continuous, positive-defined variable $Y$, the meteorological observable of interest, here the wind speed in one specific location and  given look-ahead forecast time.\\
The EMOS method uses a parametric distribution of the following general form:
\begin{equation}
Y \mid X_1, \cdots, X_M \thicksim f(Y \mid X_1, \cdots, X_M)
\label{eq:MOS}
\end{equation}
where the left-hand side denotes the fact that the distribution is conditional on the ensemble member forecasts.\\
\cite{gneiting2006calibrated} proposed the truncated normal distribution (TN) as a model for the wind speed conditional distribution $f$.
An alternative distribution,  the log-normal (LN), has been proposed by \cite{baran2015log} even if the Authors found that the TN-LN mixture model outperforms the traditional TN.
The gamma distribution was finally suggested by \cite{scheuerer2015probabilistic}. All these options, and other variants, have been extensively discussed and compared  by
\cite{wilks2018univariate}.\\
Among all possible choices, in the present study we selected the gamma distribution,  $\mathcal{G}(\mu, \sigma^2)$,
as  recently exploited by \cite{casciaro2021comparing} with excellent results. This distribution is fully described 
by two parameters: the shape parameter $k$ and the scale parameter $\theta$. Their expressions in terms of the mean $\mu$ and variance $\sigma^2$ read: $k = \mu^2 / \sigma^2$ and $\theta = \sigma^2 / \mu$ with $\mu$ and $\sigma^2$ given by:
\begin{equation}
\mu = a+b_1 X_1 + \cdots + b_M X_M
\label{eq:EMOS_mu}
\end{equation}

\begin{equation}
\sigma^2 = c+dS^2 .
\label{eq:EMOS_sigma}
\end{equation}
The coefficients $a, b_1, \cdots , b_M, c, d$ are non-negative parameters and $S^2$ is the  variance of the EPS, a quantitative measure of the ensemble spread.\\
\cite{gneiting2005calibrated} proposed a strategy based on the minimization of the Continuous Ranked Probability Score (CRPS) to determine the EMOS$_0$ free parameters \citep{hersbach2000decomposition}. In plain terms, the CRPS is defined as
\begin{equation}
CRPS(F,Y) = \int_{-\infty}^{\infty} [F(x) - H(x-Y)]^2 dx
\label{eq:crps_generale}
\end{equation}
where $F$ is the cumulative probability function associated to the gamma density function $\mathcal{G}$, $Y$ is the observation, and $H$ is the Heaviside function, which returns 0 when $x < Y$ and 1 otherwise. \\
A closed form for the CRPS for the gamma distribution has been obtained by
\cite{scheuerer2015probabilistic} making the minimization procedure easy and fast. For an observation-forecast pair ($Y, \mathbf{X}$) it reads:
\begin{equation}
crps =  Y\left[2P\left(k, \frac{Y}{\theta}\right)-1\right]-k \theta\left[2P\left(k+1, \frac{Y}{\theta}\right)-1\right]-\frac{\theta}{\beta\left(\frac{1}{2},k\right)}
\label{eq:crps}
\end{equation}
with $Y$ being the observation, $P$ the incomplete gamma function \citep{abramowitz1948handbook}, and $\beta$ the beta function. The forecast vector $\mathbf{X} = (X_1, \cdots, X_M)$ comes into the expression (\ref{eq:crps}) via the parameters $k$ and $\theta$.
The quantity to be minimized in a training set where both observations and forecasts are available is:
\begin{equation}
CRPS = \frac{1}{N}\sum_{i = 1}^N crps(\mathbf{X}_i, Y_i)
\label{eq:CRPS_tot}
\end{equation}
with $i$ denoting the i-th observation-forecast pair and $N$  is the total number of pairs in the training set, here corresponding to the whole  year 2018..\\
The CRPS combines calibration and informativeness in one index, thus 
allowing  the evaluation of predictive performance 
that is based on the paradigm of maximizing the sharpness of the predictive distributions subject to calibration \citep{gneiting2007probabilistic}.

\subsection{Accounting for non-linearities and small-scale dynamic effects: an evolution of the standard EMOS strategy}
\label{sec:EMOS+}
Very recently \citep{casciaro2021comparing} proposed a novel EMOS strategy and tested it against field measurements. The new strategy turned out to be largely superior with respect to the standard EMOS. The main strengths of the approach, baptized EMOS$_{+4r}$ by \cite{casciaro2021comparing}, is that nonlinear features can be easily and economically accounted by conditioning meteorological variables. The strategy also deals with the issue related to the model grid-point best representing weather conditions observed at the ground station. This issue, particularly severe in regions having relevant small-scale orographic variations, arises one to the  coarse spatial resolution of the EPS (as well as of any general circulation model).\\
To summarize the main idea of the method, let us consider $X_i^j$ the i-th ensemble member forecast on the j-th model grid-point ($j = 1,\cdots, 4$ spans over the nearest model grid-points to the ground station) and $S^{2\:j}$ the members variance on the j-th model grid-point. We also denote by $Z_1, \cdots, Z_q$ the $q$ categorical variables (e.g.,\ the wind direction, the hour of the day, the height of the boundary layer, among others reported by \cite{casciaro2021comparing}) expected to be useful to disentangle the EPS forecast error.\\
The EMOS$_{+4r}$ predictive distribution is  exactly as in the standard EMOS$_0$ apart the key fact that now the free parameters  are best-fitted for each combination of classes’ levels, via a training set, by minimizing the CRPS.\\
In plain terms, mean and variance of the predictive distribution are now given by:
\begin{equation}
  \mu   = a(Z_1, \cdots, Z_q) + \sum_{i=1,j=1}^{M,4} b_{ij}(Z_1, \cdots, Z_q) X_i^j 
\label{eq:EMOS_mu_cond}
\end{equation}
\begin{equation}
  \sigma^2   = c(Z_1, \cdots, Z_q) + \sum_{j=1}^{4} d_{j}(Z_1, \cdots, Z_q) S^{2 \: j} 
\label{eq:EMOS_sigma_cond}
\end{equation}
where $j$ spans  over  the  4  model  grid  points  around  the  station.\\
The last step of the strategy is to perform a final EMOS$_0$ downstream of previous calibration steps with the aim of allowing a synchronization of the forecast to the current climate trend. This aim is achieved in terms of a rolling training over the past 40 days without conditioning.

\section{The dynamic calibration}
\label{The dynamic calibration strategy}
The main strengths of the EMOS$_{+4r}$ strategy can be summarized as follow: i) it accounts for nonlinear dependencies between 
predictands and both predictors and other weather observables used as conditioning variables; ii) it deals with the issue of the model grid-point selection best representing the weather at the ground station; iii) it allows the calibrated forecast to be aligned with the current weather climate via a rolling training.\\ These remarkable strengths  are accompanied by the weak point of being unable to ingest real-time weather information. These these could be used latter on to extrapolate ahead in time the model error
which can be quantified at the present time if  observations are made available outside of the synoptic hours. Once the model error is extrapolated to the future via some suitable data-driven statistical methods, it can be exploited to further correct forecast error in the nearest future. How much the correction will be effective far from the hour at which the real-time observation is available is an issue to be addressed in the following sections.
\subsection{The proposed algorithm}
\label{sec:algo}
This section illustrates how the potential benefit of real-time observations can be optimally integrated with the EMOS$_{+4r}$ strategy giving birth to a new dynamic calibration strategy which will be also tested against different alternative approaches.

\begin{figure}[h!]
\includegraphics[width=0.6\textwidth]{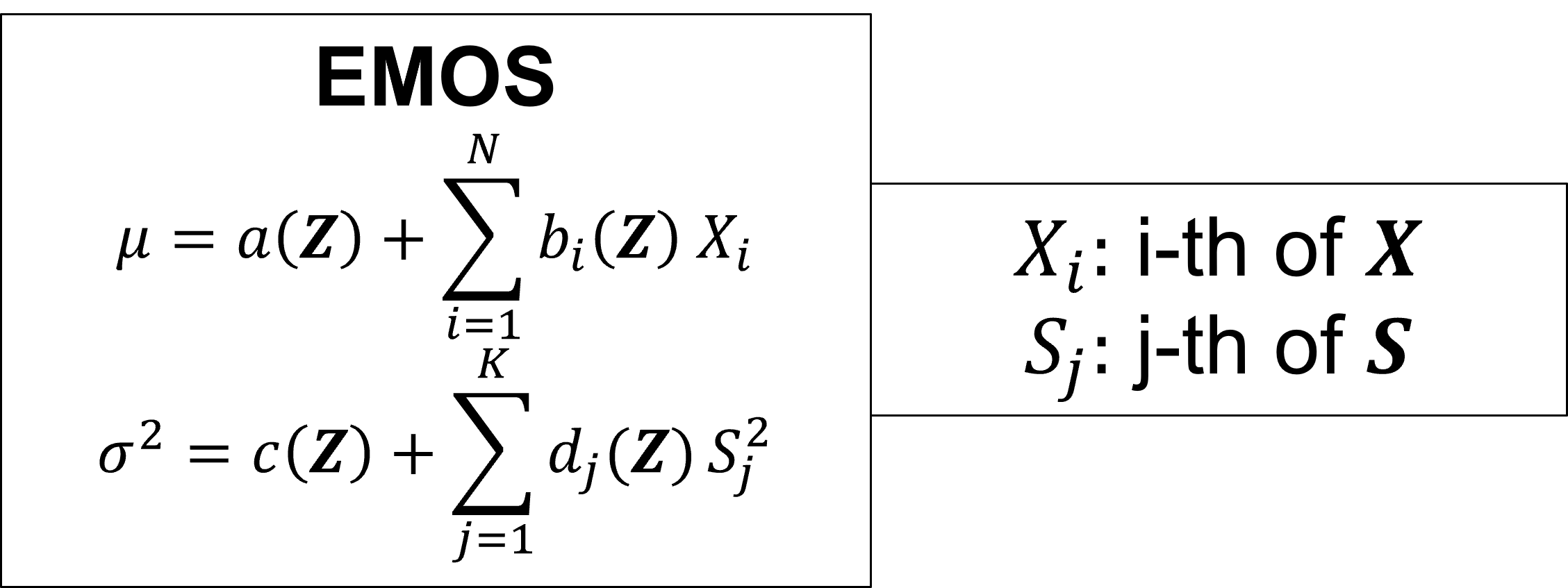}
\centering
\caption{Schematic view of the EMOS kernel of the dynamic calibration: $\mathbf{Z} = (Z_1, \cdots, Z_q)$ are $q$ categorical conditioning variables entering in the model parameters $a,b,c$, and $d$; $\mathbf{X} = (X_1, \cdots, X_N)$ are $N$ input predictors for the mean, $\mu$, of the EMOS predictive  distribution; $\mathbf{S} = (S_1, \cdots, S_K)$ are $K$ input predictors for the variance, $\sigma$, of the EMOS predictive distribution.}  
\label{fig:EMOS_generale}
\end{figure}

\begin{figure}[h!]
\includegraphics[width=0.6\textwidth]{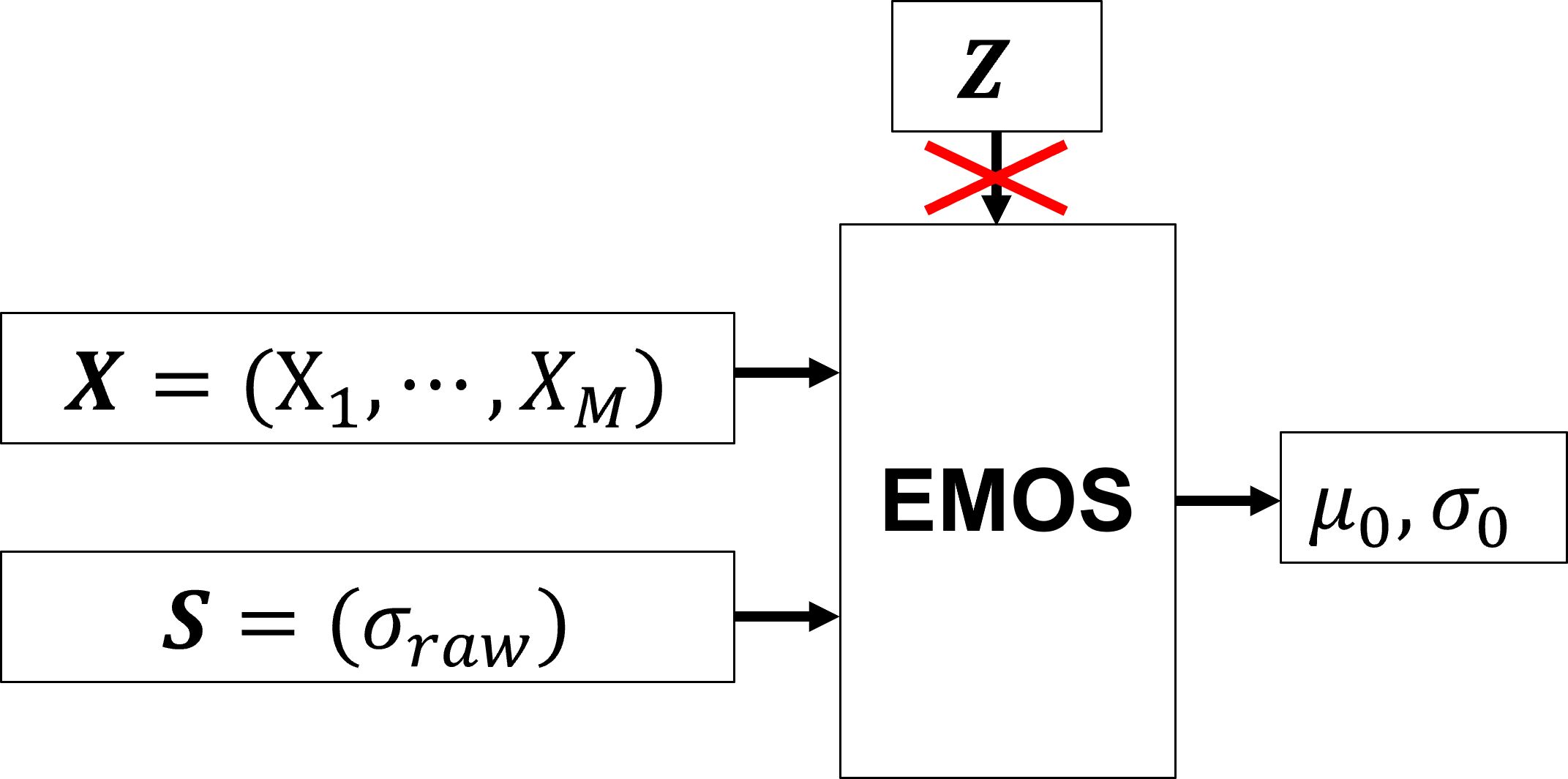}
\centering
\caption{Schematic representation of the Step 0 of our dynamic calibration in one of the 4 nearest model grid-points to a station. Input: $\mathbf{X} = (X_1, \cdots,X_M)$ are the $M$ members of the ensemble forecasts; $\mathbf{S} = (\sigma_{raw})$ is the square root of the variance of the ensemble member forecasts (K=1). Output: $\mu_0$ and $\sigma_0$ the parameters of the calibrated predictive distribution.}
\label{fig:EMOS0}
\end{figure}

\begin{figure}[h!]
\includegraphics[width=0.8\textwidth]{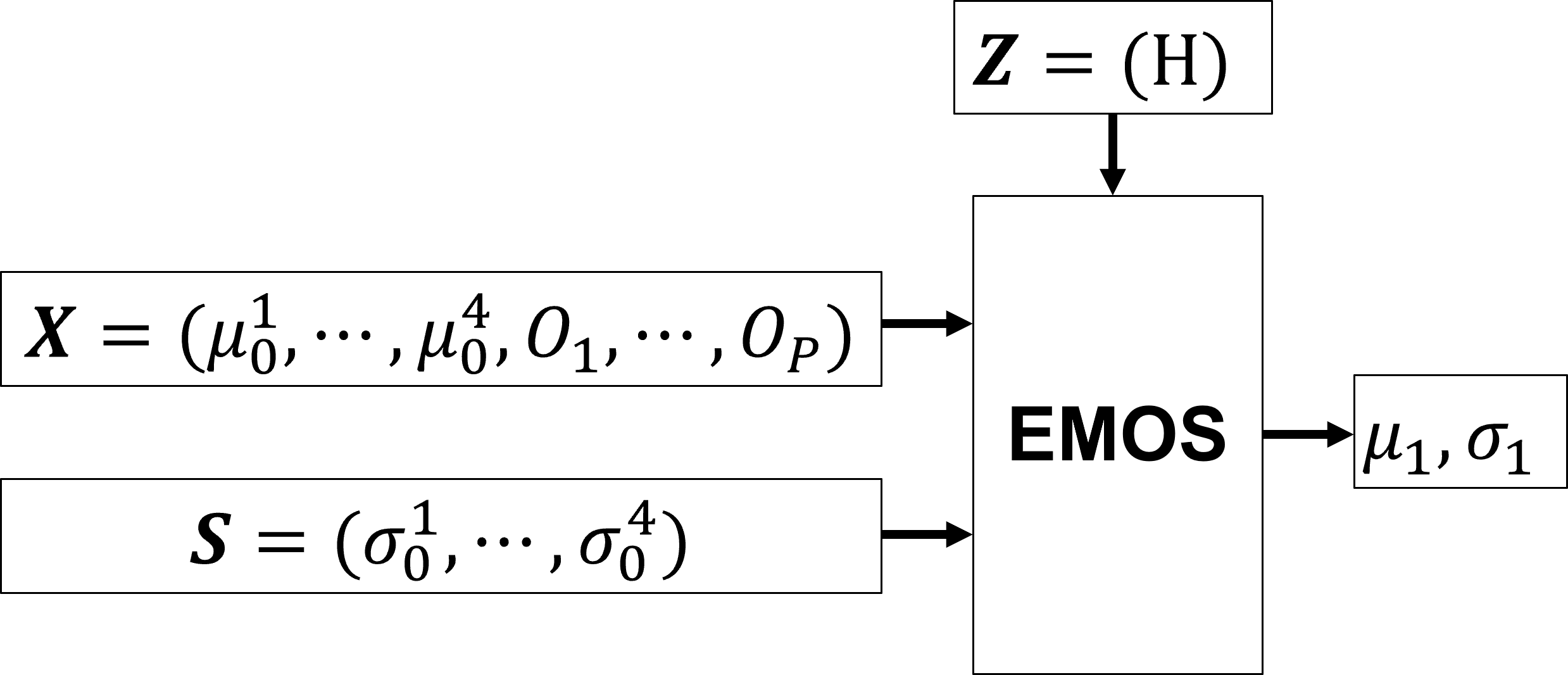}
\centering
\caption{Schematic representation of Step 1 of our dynamic calibration involving the information coming from Step 0.
Input: $\mathbf{X} = (\mu_0^1, \cdots, \mu_0^4, O_1, \cdots , O_P)$, where $\mu_0^1, \cdots, \mu_0^4$ come from Step 0 iterated for the four closest-to-the-station model grid-points,
and $(O_1, \cdots , O_P)$ are $P$ variables related to station observations available at hour $h$;
$\mathbf{S} = (\sigma_0^1, \cdots, \sigma_0^4)$ comes from Step 0. Conditioning variables: $\mathbf{Z} = (H)$ is the hour of the day (hourly step within the 24 hours).
Output: $\mu_1$ and $\sigma_1$ the calibrated parameters of the Step 1 predictive distribution.}
\label{fig:EMOS_hd}
\end{figure}

\begin{figure}[h!]
\includegraphics[width=0.8\textwidth]{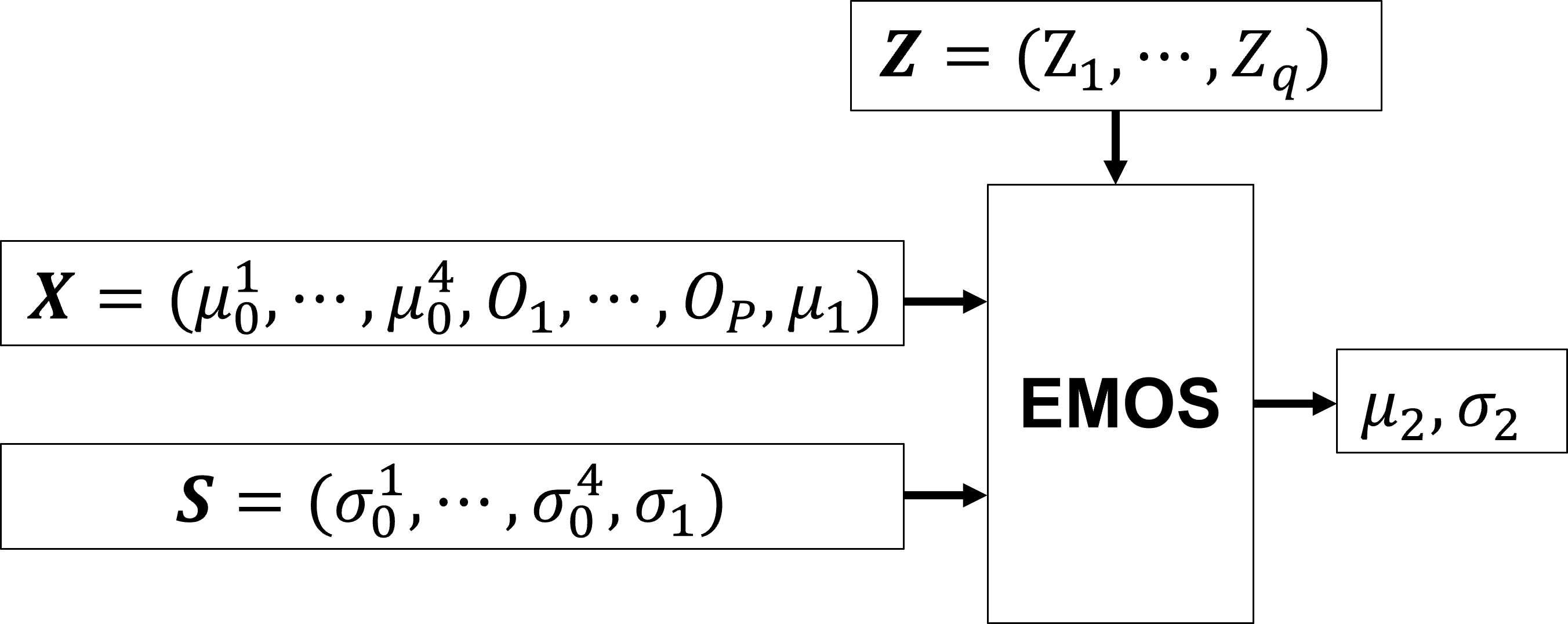}
\centering
\caption{Schematic representation of Step 2 of our dynamic calibration.
Input: $\mathbf{X} = (\mu_{0}^1, \cdots, \mu_{0}^4, O_1, \cdots , O_P, \mu_{1})$ where $\mu_0^i$ ($i=1,\cdots , 4$) comes from Step 0,
$(O_1, \cdots , O_P)$ are $P$ station observations available at hour $h$;
  $\mu_{1}$ comes from Step 1;
  $\mathbf{S} = (\sigma_0^1, \cdots, \sigma_0^4, \sigma_{1})$ where $\sigma_0^i$ comes from Step 0,
  $\sigma_{1}$ comes from Step 1.
  Conditioning variables: $\mathbf{Z} = (Z_1, \cdots, Z_q)$,
  selected following the same procedure as in \cite{casciaro2021comparing}.
  Output: $\mu_{2}$ and $\sigma_{2}$, the Step 2 calibrated parameters of the predictive gamma distribution.}
\label{fig:EMOS_+4d}
\end{figure}

\begin{figure}[h!]
\includegraphics[width=0.55\textwidth]{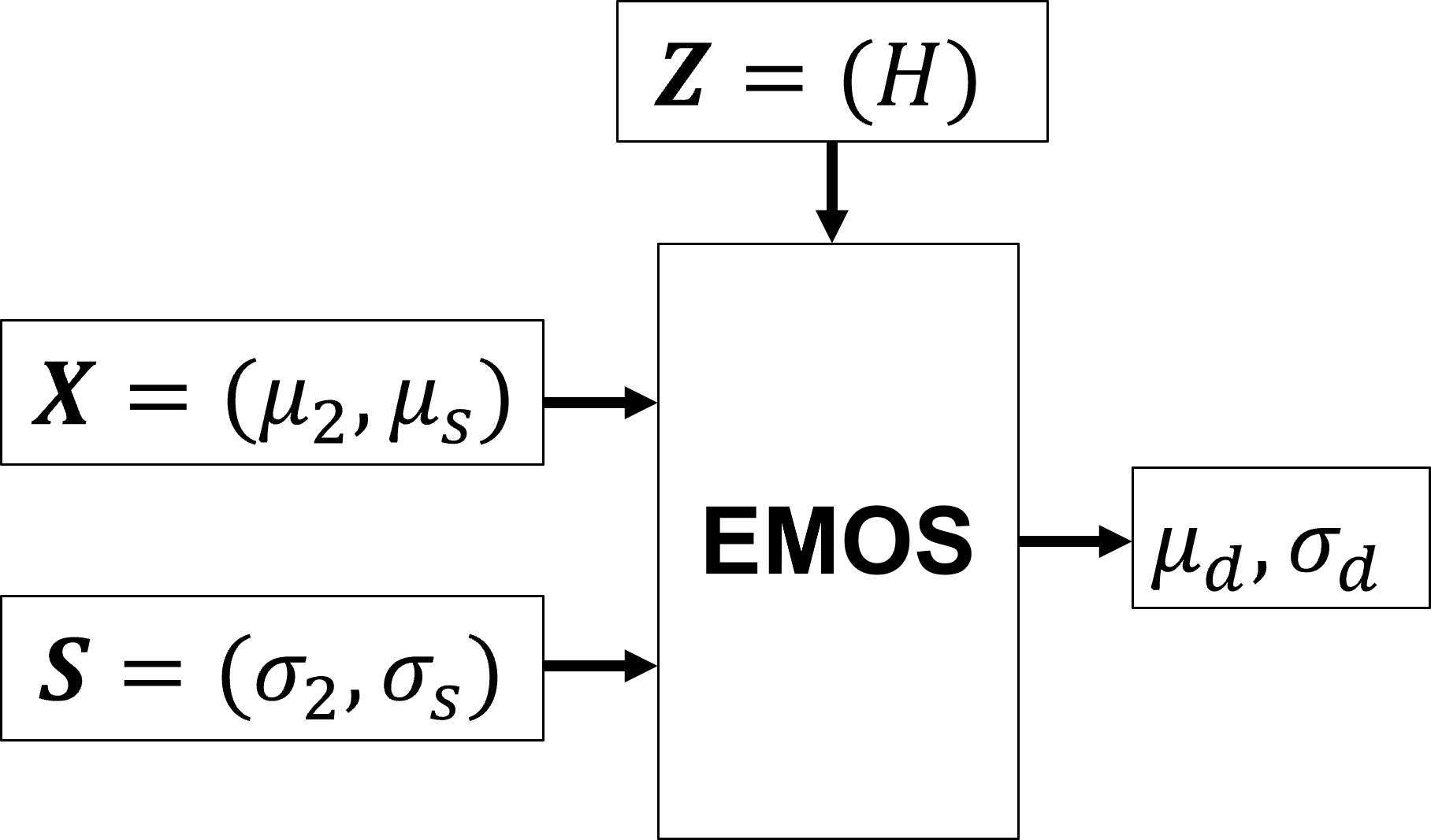}
\centering
\caption{Schematic representation of Step 3, the last step of our dynamic calibration.
  Input: $\mathbf{X} = (\mu_{2}, \mu_{s})$ and $\mathbf{S} = (\sigma_{2}, \sigma_{s})$
  where $\mu_{s}$ and $\sigma_{s}$ come from the EMOS$_{+4r}$ static calibration by \cite{casciaro2021comparing} and reviewed in Sec.\ \ref{sec:EMOS+}. The parameters  $\mu_{2}$ and $\sigma_{2}$ come from Step 2 of our calibration.
  Conditioning variables:   $\mathbf{Z} = (H)$ as in Fig.\ \ref{fig:EMOS_hd}.
Output: $\mu_{d}$ and $\sigma_{d}$ the parameters of the final calibrated gamma predictive probability density function.}
\label{fig:EMOS_+4sd}
\end{figure}

In order to ensure greater clarity and better understanding, we report in Figure \ref{fig:EMOS_generale} the kernel of the calibration, i.e.\ a general EMOS having N input predictors for the ensemble mean, K input predictors for the ensemble variance, and q conditioning variables. The values of N, q, and K will vary from case to case in the different steps of the calibration we will show in the following.
%
%
The first step (here denoted by Step 0) is to apply the EMOS kernel
to each of the 4 nearest model grid-points to the SYNOP station without using conditioning variables. This step is schematically depicted in Figure \ref{fig:EMOS0} for a representative model grid-point.
By applying it to each of the four model grid-points, we end up with 
$\mu_0^1, \cdots, \mu_0^4; \sigma^1_0, \cdots, \sigma^4_0$, with the upper indices being associate to the model grid-point among the four nearest to the ground station. The purpose of this first step is simply to reduce the forecast BIAS while maintaining unchanged the correlation between forecasts and observations.
The resulting calibrated parameters from Step 0 are used as input to the second step of our calibration, Step 1.\\
New players $(O_1,\cdots , O_P)$ come into Step 1 and involve $P$ observed data available at a reference hour, say $h$ (i.e.\ $h$ is the last hour at which observed data are available and usable). In way of example, $P=2$ with $O_1$ being the persistence built from the wind speed known at hour $h$, and $O_2$ is the diurnal-cycle-based persistence, i.e.\ a forecast built in terms of the observations occurred in the past 24 hours. 
Figure \ref{fig:EMOS_hd} summarizes how observations are combined with the results from Step 0 carried out for the four closest-to-the-station model grid-points. 
From Step 1 we end up with  the parameters $\mu_{1}, \sigma_1$ of the predictive distribution. This latter now becomes a function not only of the ensemble members but also of suitable observations.\\
Having used the hour of the day as conditioning variables, the information on how far one is from the last available observation (at hour $h$) indirectly comes into the model.\\
We now need to join all previous steps in terms of suitable conditioning variables. This is the Step 2 of our calibration as summarized in  Fig.\ \ref{fig:EMOS_+4d}.
The result of this last step of calibration is the calibrated predictive probability gamma  function, $\mathcal{G}(\mu_{2}, \sigma_{2})$, which selects the best model grid-point performance, disentangles the model error in terms of suitable conditioning variables and, finally, synchronizes  the forecast with the most recent observed wind speed at the station. We however found (see Sec.\ \ref{Sec:eval-step} for a quantitative analysis) that one more step is needed in order to extend  ahead in time as much as possible  the benefit of real-time observations. This final step (Step 3) is sketched  in Figure \ref{fig:EMOS_+4sd}. It  simply joins the static calibration EMOS$_{+4r}$ by \cite{casciaro2021comparing}, with parameters $\mu_{s}, \sigma_{s}$, (where the subscript 's' is meant to stress out that such parameters are from a `static' calibration) with those coming from Step 2. The final result is the gamma probability density function,  $\mathcal{G}(\mu_{d}, \sigma_{d})$, with $\mu_{d}$ and $ \sigma_{d}$ coming from Step 3 of Fig.\ \ref{fig:EMOS_+4sd}.\\

\subsection{Quantifying the importance of the different steps of calibration}
\label{Sec:eval-step}
\begin{figure}[h!]
\includegraphics[width=\textwidth]{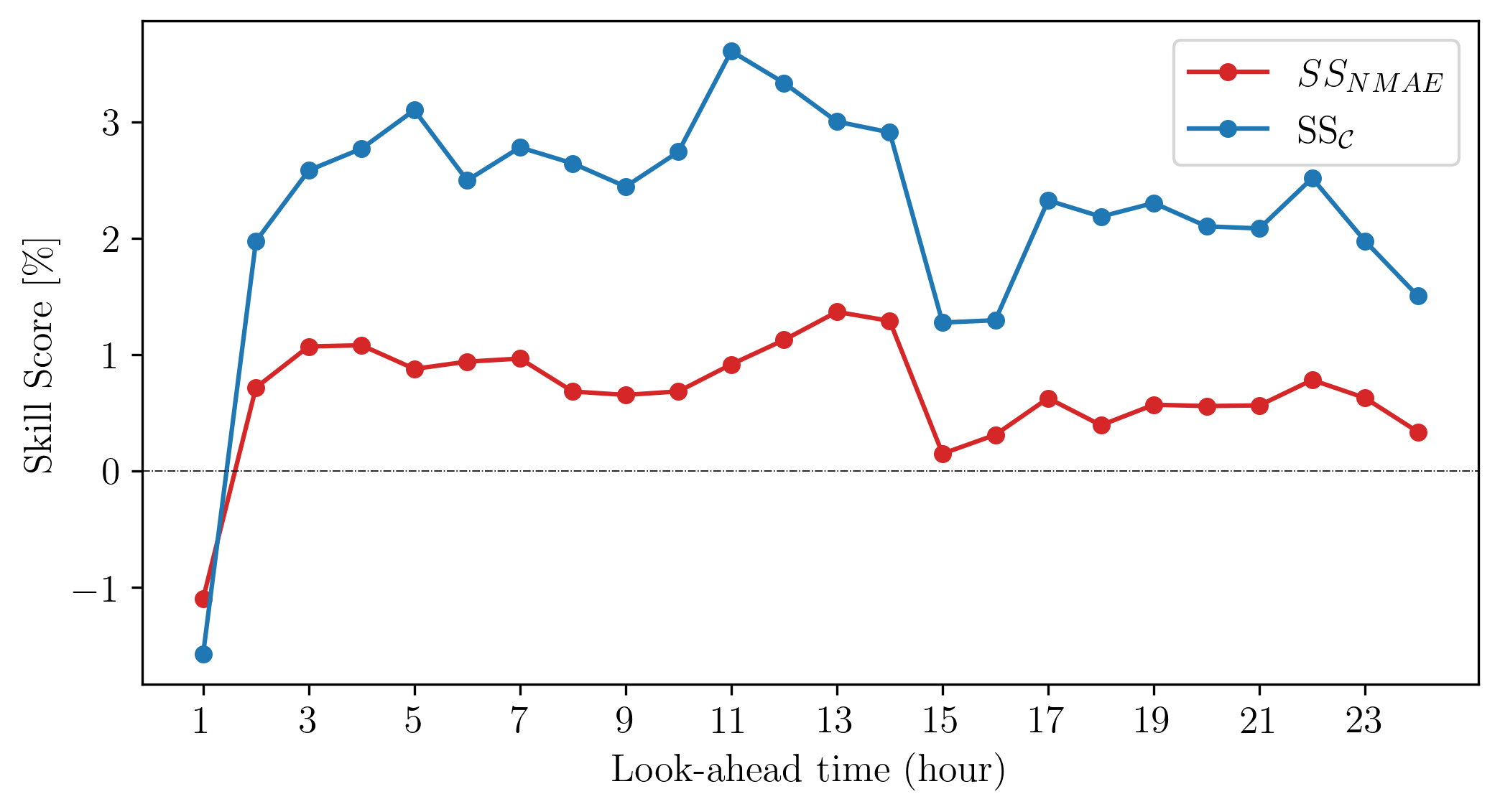}
\centering
\caption{Mean skill score (of the skill scores computed at each station) for both the NMAE index and the correlation coefficient, as a function of the look-ahead time, for the complete dynamic calibration (i.e.\ the one starting from Step 0) where the observed data at  h = 9 UTC are used. The reference calibration against which the skill scores have been  calculated is the same complete dynamic calibration which starts from Step 1 using the $EMOS_{+4r}$ mean and variance as input.  The abscissa refers to the 24-hour forecast horizon starting from the hour h = 9 UTC  where the observations are available, so that look-ahead time equal to 1 corresponds the forecast time at the hour h+1 UTC, and so on for the other abscissa values.}
\label{fig:step0_noE0}
\end{figure}

\begin{figure}[h!]
\includegraphics[width=\textwidth]{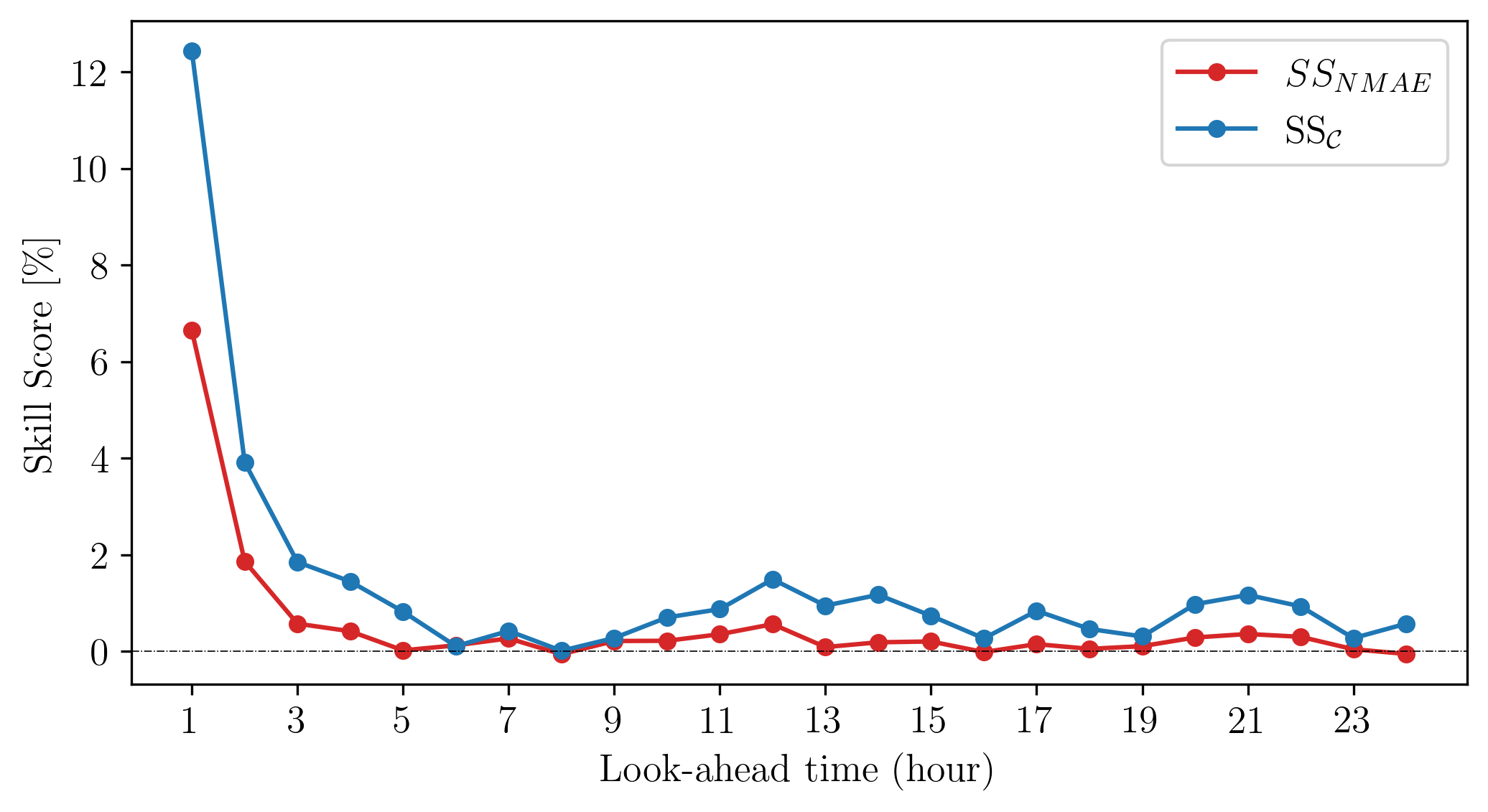}
\centering
\caption{As in Fig.\ \ref{fig:step0_noE0} but now the reference calibration is the one carried out
  in terms of our dynamic calibration without Step 2.}
\label{fig:nostep2}
\end{figure}

\begin{figure}[h!]
\includegraphics[width=\textwidth]{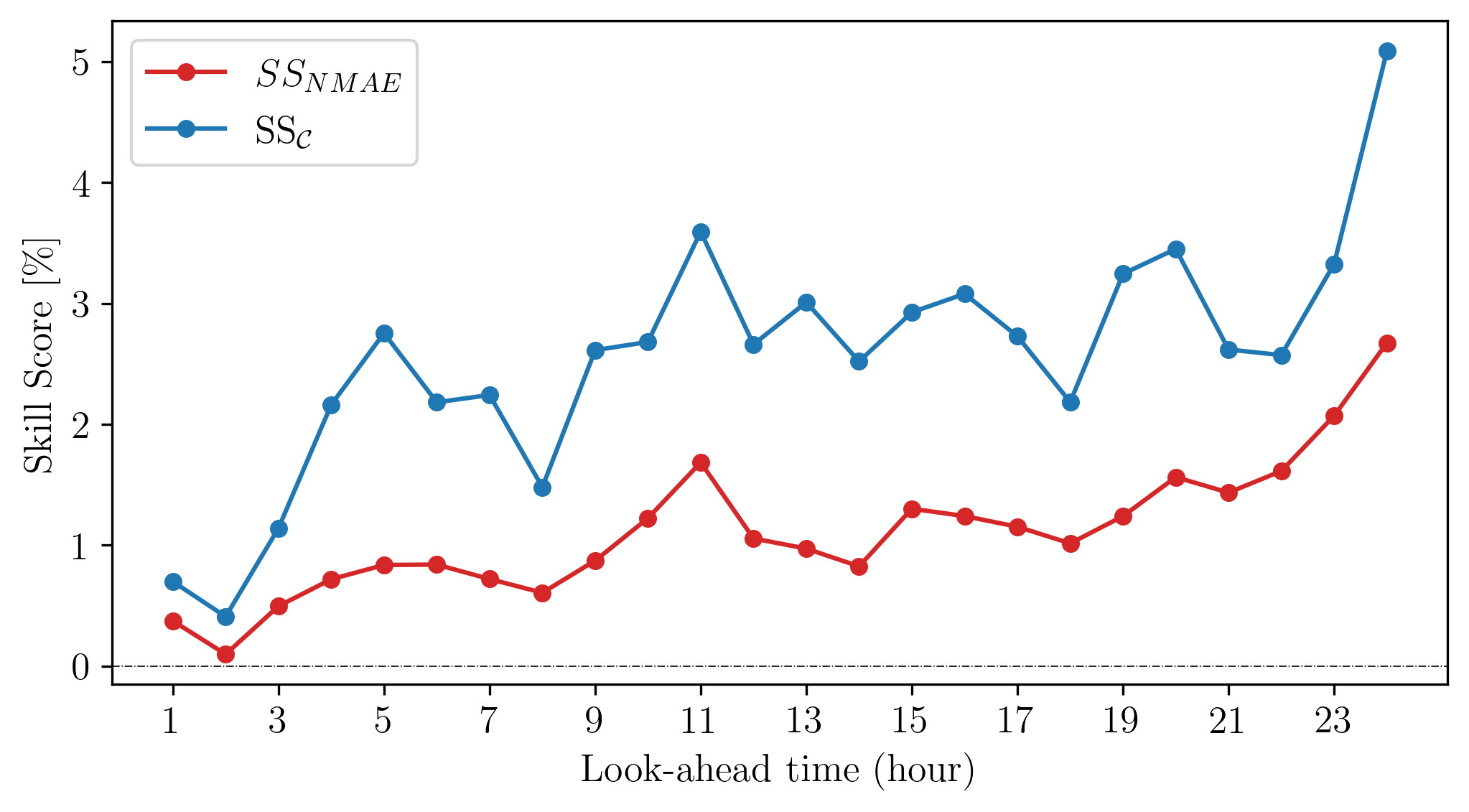}
\centering
\caption{As in Fig.\ \ref{fig:step0_noE0} but now the reference calibration is the one carried out
  in terms of our dynamic calibration without Step 3.}
\label{fig:nostep3}
\end{figure}

Let us start to quantify the added value  brought by single steps of our dynamic
calibration by assuming that observed data are available at h = 9 UTC.
Here $P=1$ and $O_1$ is the persistence built from the wind speed known at hour $h$.
In order to quantify the benefit brought by  Step 0 of our calibration,
we compare in Fig.\ \ref{fig:step0_noE0} the results of our dynamic calibration obtained with and without Step 0, respectively. This is done in terms of the skill score (see \ref{App:stat}) where the skill of the calibrated forecast from the complete calibration (i.e.\ starting from Step 0)  is comparatively quantified against the calibrated forecast from the dynamic calibration now starting directly from Step 1 using the $EMOS_{+4r}$ mean and variance as input. As one can see from the figure, apart the first hour of forecast, there is a clear added value brought by Step 0, as quantified both in terms of the NMAE index and in terms of the correlation coefficient.  
In Fig.\ \ref{fig:nostep2} the importance of Step 2 is quantified similarly to what we did in
Fig.\ \ref{fig:step0_noE0}:  for both the NMAE index and the correlation coefficient, the skill score of the complete dynamic calibration is computed taking the forecast calibrated from the dynamic calibration without Step 2 as a reference.
The results from  Fig.\ \ref{fig:nostep2} clearly indicate the added value brought by Step 2, especially in the first 5 hours of forecast.\\
We conclude by quantifying the role of Step 3. This is done in Fig.\ \ref{fig:nostep3} where the reference calibrated forecast is the one coming from the dynamic calibration now without using Step 3. Also in this case the importance of the selected Step 3 is evident.

\subsection{Assessing the dynamic calibration via statistical indices}
\label{Sec:assess}
\begin{figure}[h!]
\includegraphics[width=\textwidth]{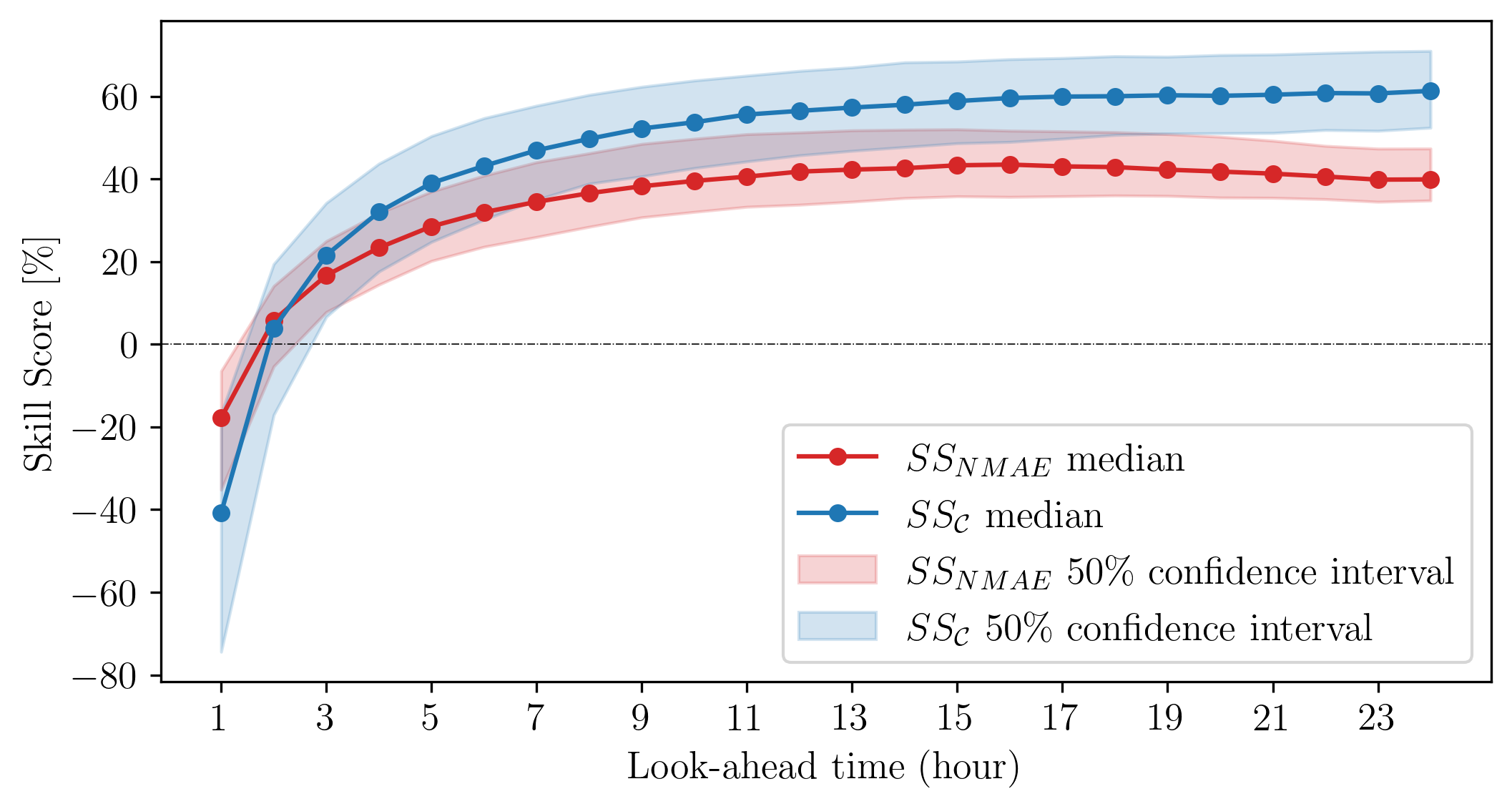}
\centering
\caption{Skill score of NMAE and correlation coefficient of the $EMOS_{+4r}$ static calibrated forecast using the persistence at time h as a reference forecast. Continuous lines: median of the skill scores computed by considering all stations and all hours h within the first 24 hours from 00 UTC. Shaded areas represent the 50$\%$ confidence interval of the skill scores, accounting for the variability of the latter among stations and while varying the hour h at ourly steps between 00 UTC and  24 UTC of the same day.}
\label{fig:st-vs-pr-det}
\end{figure}

\begin{figure}[h!]
\includegraphics[width=\textwidth]{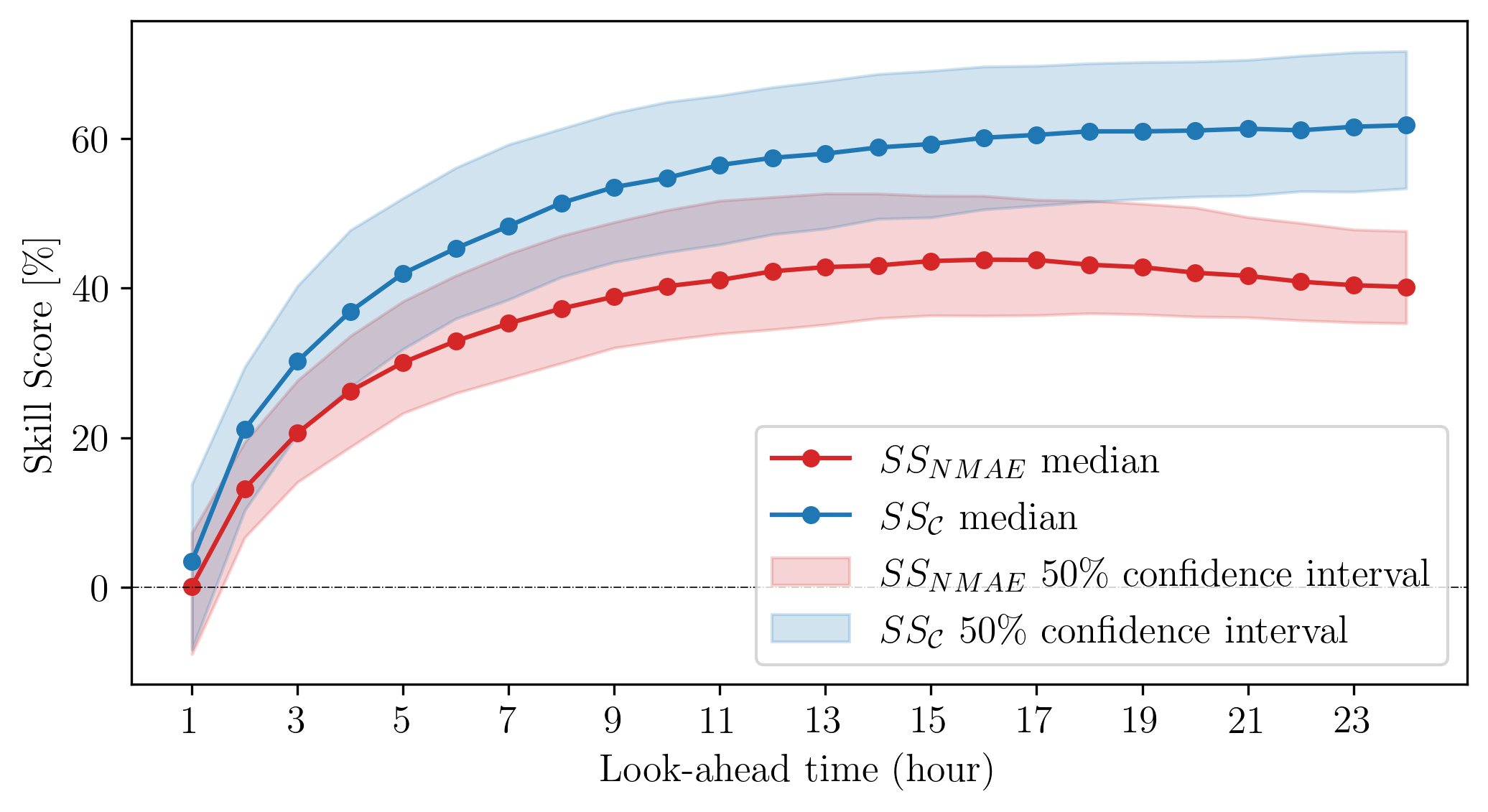}
\centering
\caption{As in Fig.\ \ref{fig:st-vs-pr-det} but now the skill of our dynamic calibration 
is assesed against persistence.}
\label{fig:dn-vs-pr-det}
\end{figure}

\begin{figure}[h!]
\includegraphics[width=\textwidth]{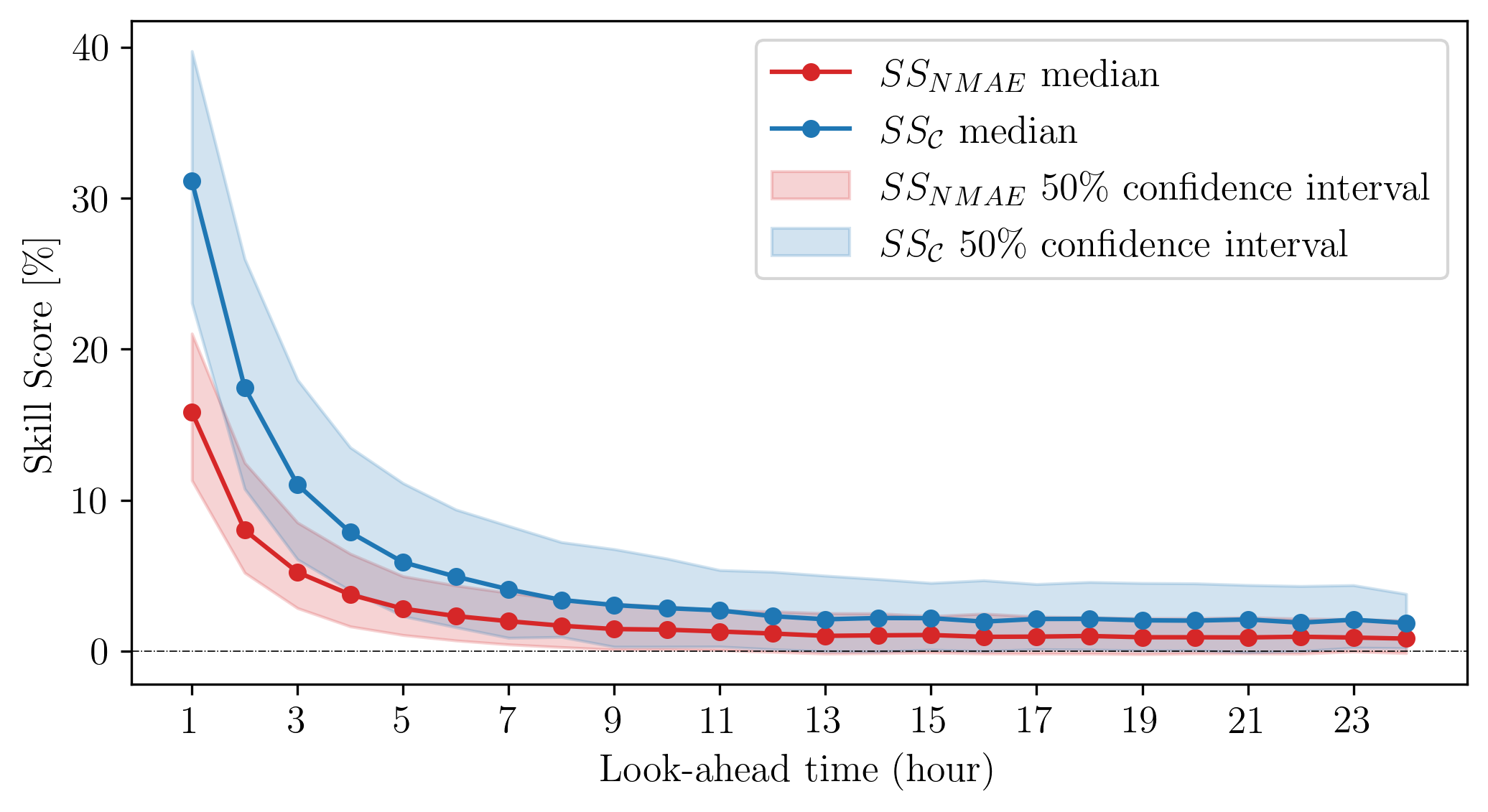}
\centering
\caption{As in Fig.\  \ref{fig:st-vs-pr-det} but now the skill of our dynamic calibration 
is assesed against the static $EMOS_{+4r}$ calibration.}
\label{fig:dn-vs-st-det}
\end{figure}

Having justified the added value brought by single steps of our dynamic calibration,
we now pass to assess the performaces of the dynamic calibration against different forecasts,
both static, based on calibrated forecast from NWP models, and statistical, built solely in terms of past observed data, using persistence (i.e.\ the simplest way to build a forecast from a given observation)  and a ML-based prediction.
Here we consider the simplest way to produce a forecast from observed data: the
 wind speed known at hour $h$ is maintained for all considered look-ahead forecast horizons. This is thus a prediction based on persistence whose values are encoded in the variable $O_1$  in Step 1 and 2  of our dynamic calibration with $P=1$. \\
Let us start by emphasizing the weakness of the static calibration $EMOS_{+4r}$ in the first forecast hours, a fact that motived us to propose a dynamic calibration. To do that, let us imagine to have real-time data available at the hour h (h belongs to the first 24 hours from the 00 UTC) and to have at our disposal the $EMOS_{+4r}$ static calibration starting from 00 UTC with a forecast horizons of 48 hours.  The $EMOS_{+4r}$ skill score (for both NMAE and correlation coefficient
${\cal C}$) is reported in Fig.\ \ref{fig:st-vs-pr-det} where the abscissa refers to the 24-hour forecast horizon starting from the hour h  where the observations are available (so that an abscissa equal to one corresponds to the forecast time at the hour h+1). Such observation is not ingested by the static $EMOS_{+4r}$ calibration but is used here to build a forecast based on persistence which serves as reference forecast to evaluate the $EMOS_{+4r}$ skill score of Fig.\ \ref{fig:st-vs-pr-det}.
The obtained skill scores are computed averaging over all stations and over all hours h in the first 24 hours starting from 00 UTC.
The shaded areas represent the 50 $\%$ confidence interval around the median thus providing us an idea on how
the skill score varies among the different stations and by varying the hour h at which data are available.
As one can clearly see from Fig.\ \ref{fig:st-vs-pr-det} persistence overcomes $EMOS_{+4r}$ in the first forecast hour while rapidly deteriorating from the second hour onwards. This holds true  in terms of both NMAE and correlation coefficient.
The performance of our dynamic calibration is shown in Fig.\ \ref{fig:dn-vs-pr-det} where
the associated  skill scores are shown by taking the persistence as a reference forecast. 
As one can see, our strategy overcomes persistence from the first hour onwards while maintaining the
quality of the $EMOS_{+4r}$ static calibration already detectable  from Fig.\ \ref{fig:st-vs-pr-det}
for sufficiently long forecast horizons.
This last remark can be clearly detected from Fig.\ \ref{fig:dn-vs-st-det} where the skill score of the dynamic calibration is now presented taking the static calibration from the $EMOS_{+4r}$ as a reference. A remarkable added value brought from the dynamic calibration clearly emerges in the first 6-7 hours, progressively reducing as the forecast horizon increases. Interestingly, the benefit carried by the ingestion of the observed data never disappears, even for the farest look-ahead forecast times.\\
Let us conclude the assessment in terms of point error indices
by evaluating the skill of our dynamic calibration against the
prediction built from a ML-based algorithm detailed in \ref{App:ML}.
\begin{figure}[h!]
\includegraphics[width=\textwidth]{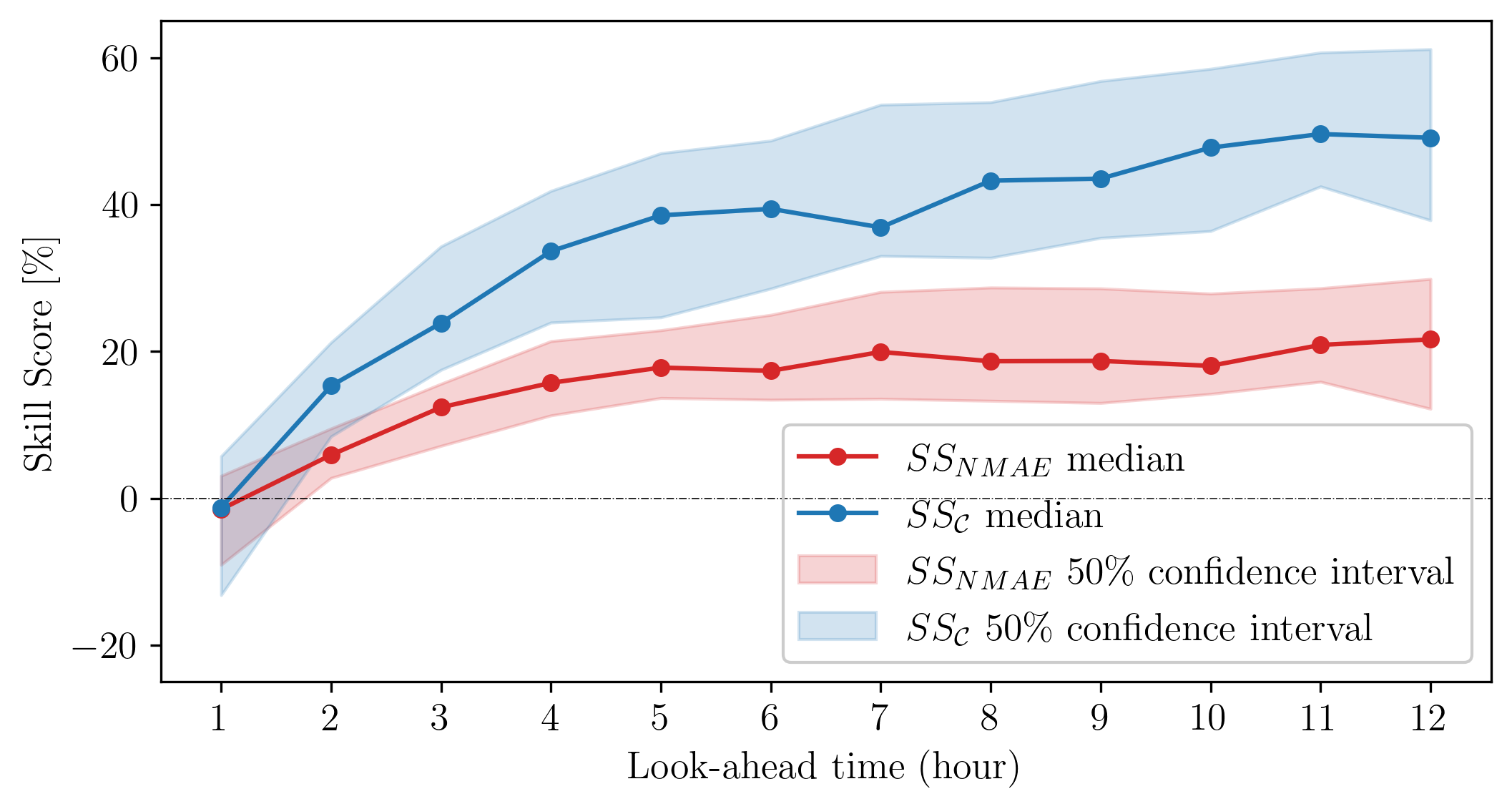}
\centering
\caption{As in Fig.\ \ref{fig:st-vs-pr-det} but now the skill of our dynamic calibration 
is assessed against the ML observation-driven predictions. Also note that in this comparison we have confined the attention to the sole case corresponding to h = 9 UTC with a forecast horizon of 12 hours. The $50\%$ confidence interval thus gives us an idea on the variability of the skill score from station to station.}
\label{fig:dn-vs-krr}
\end{figure}
This latter only uses observed data available at hour h = 9 UTC
with a forecast horizon of 12 hours. Results are reported in Fig.\ \ref{fig:dn-vs-krr} and show how the dynamic calibration outperforms the observation-driven ML-based prediction.

\begin{figure}[h!]
\includegraphics[width=\textwidth]{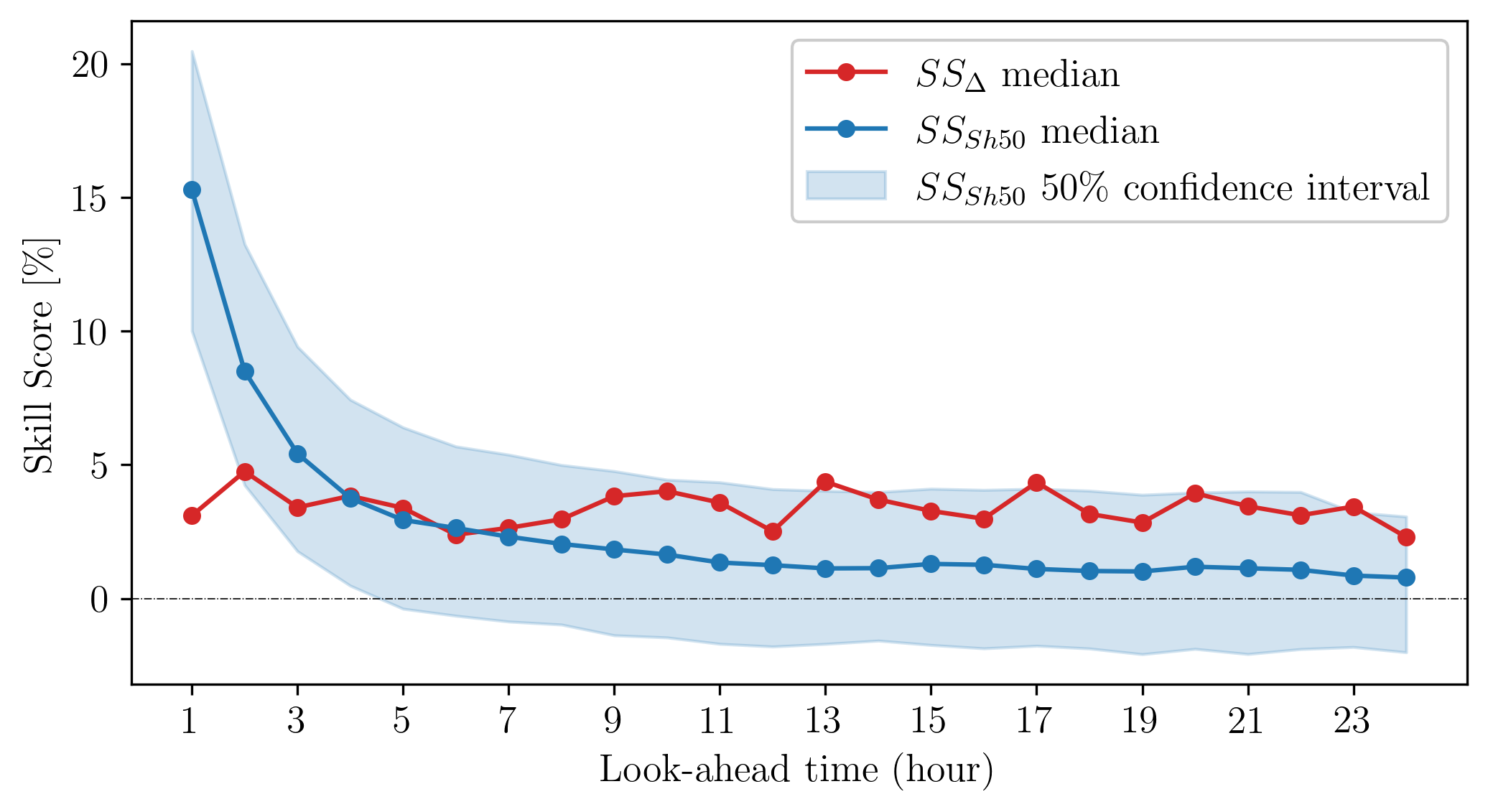}
\centering
\caption{As in Fig.\ \ref{fig:dn-vs-st-det} but now the skill scores are computed for $\Delta$ and the sharpness parameter $Sh50$. For the skill score of $\Delta$ the confidence interval (not shown) varies between -20 and 20 $\%$.}
\label{fig:dn-vs-st-prob}
\end{figure}

We have till now considered point error indices. Let us now pass to assess the predictive
probability density function as a whole. This is done in Fig.\ \ref{fig:dn-vs-st-prob} in terms of
the so-called reliability index ($\Delta$) proposed by \cite{delle2006probabilistic}, and the sharpness $Sh50$ corresponding to the  average width of the central 50\% prediction interval
of a forecast probability distribution as
proposed by \cite{gneiting2007probabilistic}. Details are reported in \ref{App:stat}.

In Fig.\ \ref{fig:dn-vs-st-prob} the reference forecast is the static $EMOS_{+4r}$ calibration.
The increased skills of the dynamic calibration overcoming the static calibration especially in the first forecast hours is evident, thus completing our quality assessment.  \\
A short summary of the most relevant error indices considered in the present study are reported in Tab.\ \ref{tab:final} for our dynamic calibration and, for comparison, for the static calibration and persistence. Results are presented for three forecast time intervals starting from the hour h at which observations are available.
For the sake of example, the interval 1-2 hours means that the reported indices are the average
of the corresponding indices at hours h+1 and h+2. A further average is performed over all possible h between 1 and 24 hours (hourly step) and over all stations.  The table confirms all conclusions drawn in the paper on the superiority of our dynamic calibration against the static one and, even more, against persistence.  

\begin{table}[]
  \label{tab:final}
\centering
\caption{A final summary of relevant indices used to assess the skill of the calibrations.}
\begin{tabular}{c|c|c|c|c}
\cline{2-4}
                           & Persistence & \begin{tabular}[c]{@{}c@{}}Static\\ calibration\end{tabular} & \begin{tabular}[c]{@{}c@{}}Dynamic\\ calibration\end{tabular} &                                            \\ \hline
\multicolumn{1}{|c|}{NMAE} & 0.32        & 0.34                                                         & 0.30                                                          & \multicolumn{1}{c|}{\multirow{3}{*}{1-2 h}} \\ \cline{1-4}
\multicolumn{1}{|c|}{$\cal{C}$}    & 0.76        & 0.71                                                         & 0.78                                                          & \multicolumn{1}{c|}{}                      \\ \cline{1-4}
\multicolumn{1}{|c|}{Sh$_{50}$} & -           & 3.5                                                         & 3.1                                                          & \multicolumn{1}{c|}{}                      \\ \hline
\multicolumn{1}{|c|}{NMAE} & 0.43        & 0.35                                                         & 0.33                                                          & \multicolumn{1}{c|}{\multirow{3}{*}{3-4 h}} \\ \cline{1-4}
\multicolumn{1}{|c|}{$\cal{C}$}    & 0.62        & 0.71                                                         & 0.74                                                          & \multicolumn{1}{c|}{}                      \\ \cline{1-4}
\multicolumn{1}{|c|}{Sh$_{50}$} & -           & 3.5                                                         & 3.3                                                          & \multicolumn{1}{c|}{}                      \\ \hline
\multicolumn{1}{|c|}{NMAE} & 0.51        & 0.35                                                         & 0.34                                                          & \multicolumn{1}{c|}{\multirow{3}{*}{5-6 h}} \\ \cline{1-4}
\multicolumn{1}{|c|}{$\cal{C}$}    & 0.52        & 0.71                                                         & 0.72                                                          & \multicolumn{1}{c|}{}                      \\ \cline{1-4}
\multicolumn{1}{|c|}{Sh$_{50}$} & -           & 3.5                                                         & 3.4                                                          & \multicolumn{1}{c|}{}                      \\ \hline
\end{tabular}
\end{table}


\clearpage

\section{Conclusions and perspectives}
\label{Sec:conclu}
A novel EMOS strategy, baptized $EMOS_{+4r}$, has been recently proposed by
\cite{casciaro2021comparing}. This strategy differs from the standard EMOS
in that the free parameters entering the predictive EMOS probability density function are categorical functions, and  not constant as in the standard EMOS.
This allows the introduction of nonlinearities in the calibration strategy.
These categorical functions must be best-fitted for each combinantion of classes’ levels, via a training set, by minimizing the CRPS.\\
If, on one hand, the new calibration strategy brings a relevant added value with respect to the standard EMOS calibration, on the other hand persistence built in terms of an observation made available at the hour h
turns out to be more accurate at the next hour  h+1, rapidly degrading at the next look-ahead forecast times. This simple observation suggested us that persistence could have been favorably ingested  in the static $EMOS_{+4r}$ calibration with the following two main aims: i) taking advantage of the known information from observed data at hour h in between two consecutive synoptic hours; ii) maintaining, and possibly increasing, the well-established superiority of $EMOS_{+4r}$ over persistence after the hour h+1. Transforming this idea into a new dynamic calibration strategy taking advantage of real-time observed data has been the main result of the present paper.\\
At the same time, the new calibration maintains  the well-established superiority of $EMOS_{+4r}$ in the far forecast horizons over purely data-driven statistical forecasts, including complex ML-based predictions.\\
The higher quality of the wind speed forecast over forecasts by static approaches not ingesting real-time observations, and remaining frozen between two successive six-hour separated synoptic hours, 
is expected to generate a relevant added value for the wind power forecast with many useful applications for the whole wind industry. \\
To conclude with some perspectives, our new strategy paves the way to a myriad of other applications of interest for the green energy market. What we did for the wind speed can indeed be easily generalized to other green sources, including solar radiation and wave generation
\citep{lira2021future,ferrari2020optimized,besio2016wave}. The key ingredient for the strategy to be implemented is to have at disposal observed data in real/quasi-real time, a requirement getting easier every day.

\section{Acknowledgments}
\noindent
G.C. has been funded by the Italian bank foundation ``Fondazione Carige''. A.M. acknowledges the funding from the Interreg Italia-Francia Marittimo SICOMAR+ project (grant number D36C17000120006) and from the Compagnia di San Paolo (Project MINIERA No. I34I20000380007). A.L.L has been funded by the Interreg Italia-Francia Marittimo SINAPSI Project (grant number D64I18000160007). We thank the Aeronautica Militare - Servizio Meteorologico - for providing us with the SYNOP data as well as data from the EPS forecasts. Discussions with Lorenzo Rosasco, Agnese Seminara, and Alessandro Verri  are warmly acknowledged.

\appendix

\section{Statistical indices}
\label{App:stat}

The Skill Score (SS) index \citep{wilks2011statistical} is used here to make the comparison
between different calibration strategies as quantitative as possible. It assesses the performance 
of a given calibration by comparing its associated statistical error index against the one corresponding to a reference forecast. Namely, 
\begin{equation} 
SS = \frac{A - A_{ref}}{A_{opt} - A_{ref}}
\label{eq:Skill-Score}
\end{equation}
where $A$ is the value of a suitable error index associated to the calibrated forecast, $A_{ref}$ is the same as $A$ but relative to a reference forecast. Finally, $A_{opt}$ refers to the optimal
index value.  A perfect calibration yields SS=1, corresponding to the upper bound of SS. 
Values of SS smaller than one (including negative values) indicate that the calibrated forecast is less accurate than the reference one.\\
As far as the error indices are concerned ($A$ in Eq.\ (\ref{eq:Skill-Score})), here we consider  
the normalized mean absolute error (NMAE), the correlation coefficient (${\cal C}$), the so-called reliability index ($\Delta$) proposed by \cite{delle2006probabilistic}, and the sharpness $Sh50$
corresponding to the  average width of the central 50\% prediction interval
of a forecast probability distribution as
proposed by \cite{gneiting2007probabilistic}.\\
The NMAE is defined as:
\begin{equation}
NMAE = \frac{\sum_{n=1}^{N}\left|X_n - Y_n\right|}{\sum_{n=1}^{N}Y_n}
\label{eq:NMAE}
\end{equation}
where $Y_n$ is the n-th observation and $X_n$ is the corresponding n-th forecast (here corresponding to the mean of the 50 EPS ensemble) and $N$ is the number of observation-forecast pairs in a given test set.\\
The correlation coefficient, $\cal{C}$, is a measure of linear dependence between two variables and ranges from -1 to 1, with 1 representing the highest correlation and -1 representing the highest anti-correlation \citep{wilks2011statistical}. Quantitatively \citep{lee1988thirteen},
\begin{equation} 
{\cal{C}} = \frac{\sum_{n=1}^{N}(X_n-\overline{X})(Y_n-\overline{Y})}{N\sigma_X \sigma_Y}
\label{eq:Pearson}
\end{equation}
with:
\begin{equation}
\sigma_X = \sqrt{\frac{\sum_{n=1}^{N}(X_n-\overline{X})^2}{N}}
\end{equation}
\begin{equation}
\sigma_Y = \sqrt{\frac{\sum_{n=1}^{N}(Y_n-\overline{Y})^2}{N}}
\end{equation}
where $\overline{X}$ and $\overline{Y}$ are the mean values of $X$ and $Y$.\\
The goal of the probabilistic forecast, according to \cite{gneiting2007probabilistic}, is to maximize the sharpness of the predictive distribution subject to calibration. \cite{anderson1996method} and \cite{hamill1997verification} proposed the use of verification rank (VR) histograms to assess the calibration of ensemble forecasts. VR histograms show the distribution of the ranks when the ranks of the observations are pooled within the ordered ensemble forecasts.
In a calibrated ensemble, the observations and ensemble predictions should be interchangeable, resulting in a uniform VR histogram. The continuous analogue of the VR histogram is the probability integral transform (PIT) histogram \citep{dawid1984present, diebold1997evaluating, gneiting2007probabilistic}. The PIT value is determined by the value of the predictive cumulative distribution function at the verifying observation. For calibrated forecasts, the empirical cumulative distribution function of PIT values should converge to the uniform distribution.\\
The reliability index $\Delta$ was proposed by \cite{delle2006probabilistic} to quantify the deviation of VR histograms from uniformity. To quantify the deviation from uniformity in the PIT histograms, we use here the following definition of $\Delta$:

\begin{equation} 
\Delta = \sum_{i=1}^{m}\left|f_i - \frac{1}{m}\right|
\label{eq:Delta}
\end{equation}
where $m$ is the histogram  number of classes, each with a relative frequency of $1/m$, and $f_i$ is the observed relative frequency in class $i$.\\
This index ranges from 0 to $+\infty$, with the bound zero corresponding to optimality.\\
Sharpness is a property of forecasts that refers to the concentration of predictive distributions. The sharper the forecasts, the more concentrated the predictive distributions are, and the sharper the better, subject to calibration.\\
To assess sharpness we average the widths of the central 50\% prediction intervals
of forecast probability distributions at all time instants. Namely,
\begin{equation}
Sh_P = \frac{1}{T} \sum_{t = 1}^T\left[Q\left(f(t), \frac{1}{2} + \frac{P}{2}\right) - Q\left(f(t), \frac{1}{2} - \frac{P}{2}\right)\right]
\label{eq:Sharp}
\end{equation}
where $T$ is the total number of instants, $Q$ is the quantile of the forecast distribution, $f(t)$ is the predictive distribution at time $t$ and $P$ is the probability interval (here 50\%).

\section{Technical Note on Machine Learning Models used as Benchmark}
\label{App:ML}

As a further  benchmark, beside the naive Persistence Method, we use  a Machine Learning approach which is based only on time series of measured data.
In last years Machine Learning algorithms have been proliferating in the field of data-driven wind speed prediction, establishing itself as a competitive alternative to state-of-the-art statistical models in the field of time series forecasting in general \citep{parmezan2019evaluation}. Here we adopt the strategy proposed in \cite{lagomarsino2022physics}, where wind time series are used to map the forecast task into Supervised Learning problems which they solve with a nonlinear kernel method \citep{bishop2016pattern}.
In \cite{lagomarsino2022physics} they find that, depending on location, direction can be useful to increase the prediction accuracy. To exploit this information they also consider the two Cartesian components of the wind vector $u_t,v_t$ which can be easily calculated from the original signals of wind speed $s_t$ and direction $\theta_t$ as $v_t = s_t\sin{(\theta_t)}$ and $u_t = s_t\cos{(\theta_t)}$ (being the angle $\theta_t$ measured in radiants).
Following their steps we define Machine Learning models which are able to learn, for each time $t$, the relation between the future value of the wind speed at time $t+h$ and the past $\mu$ measurements (we will also refer to $\mu$ as memory), i.e. 
\begin{equation} 
\widehat{s}_{t+h}=\mathcal{F}(\eta_{t - \mu + 1}, \dots, \eta_t)
\label{eq:ML1}
\end{equation}
where $\mathcal{F}$ denotes the desired relation, $\widehat{s}_{t+h}$ our prediction at horizon $h$ and $\eta_t$ the values of the input variables at time $t$. \\
Learning function $\mathcal{F}$ can be seen as a regression problem in the context of Supervised Learning. To solve the problem, input-output couples $(\vec{x}_t,y_t)_{t=1}^n$ can be defined as

\begin{equation} 
\vec{x}_t= \big[\eta_{t - \mu + 1},\dots,\eta_t\big]\in\mathbb{R}^d
\label{eq:ML2}
\end{equation}
\begin{equation} 
y_t=s_{t+h}
\label{eq:ML3}
\end{equation}

where $n$ is the number of provided samples, $d = \mu\times k$ and $k$ is the number of variables used to define $\eta_t$. Denote by $\mat{X}\in\real{n\times d}$ (for a certain $d$) the lag matrix whose rows are $\vec{x}_t$ and $\vec{y}\in\real{n}$ be the vector of the outputs with elements $y_t$.
Linear regression assumes that future wind behavior depends linearly on its past trends: it aims to find coefficients $\vec{\beta}\in\real{d}$ such that the following error term is minimized
\begin{equation} 
\frac{1}{n}\sum_{i=1}^n \| \vec{x}_i \vec{\beta} - y_i \| ^2
=
\frac{1}{n} \| \mat{X} \vec{\beta} - \vec{y} \|^2
\label{eq:ML4}
\end{equation}

and hence the formula $\vec{x}_\mathrm{new}\vec{\beta}$ is used to infer how the wind will evolve in the future.
To allow our models to capture more complex dependencies we perform Kernel Ridge Regression (KRR) introducing a non-linear transformation of the features via the kernel function $k: \real{d}\times\real{d}\to\real{}$  \citep{murphy2012machine}. In our experiments we use the Gaussian kernel which is defined as $k(\vec{x}_i, \vec{x}_j) = e^{-\| \vec{x}_i - \vec{x}_j \| ^2 / (2\sigma^2)}$. The solution to the KRR problem yields an estimator $\widehat{f}$ which can be used for inference

\begin{equation} 
\widehat{f}(\vec{x}_{\mathrm{new}}) = k(\vec{x}_{\mathrm{new}}, \mat{X})(\mat{K} + n\lambda \mathrm{I})^{-1} \vec{y} 
\label{eq:ML5}
\end{equation}

where $\mat{K}\in\real{n\times n}$ is the kernel matrix with values $\mat{K}_{ij} = k(\vec{x}_i, \vec{x}_j)$ and $\lambda$ is a regularization parameter which ensures the problem is well-posed and must be tuned in order to optimize the predicting performance of the model.\\
Since we focus on predictions delivered daily at 9:00 a.m. then only inputs vectors whose reference times $t$ correspond to that time of day have been kept. Furthermore we aim to predict the wind speed from $1$ to $12$ hours ahead with an hourly frequency and for each horizon we consider different combinations of memory $\mu$ and input data $\eta_t$. We analyse all memories $\mu$ up to $24$ hours and two different cases for $\eta_t$. In the first case $\eta_t = s_t$, i.e.~the input data is the wind speed itself; in the second case $\eta_t = (u_t, v_t)$: the two components of the wind vector, which encodes the wind speed but also the direction. 
A reduction of the dimension of input vectors have also been evaluated performing a Principal Component Analysis on the input samples and keeping subsets of principal components with increasing importance. Finally for each pair of horizon and location we select the setup, in terms of memory and input design (either with or without dimensionality reduction), that maximises the Pearson Correlation between predictions and actually observed values of the speed.  Data from each station have been split into disjoint training and test set as described in section \ref{sec:EMOS0}. Again following \cite{lagomarsino2022physics} a static rather than a rolling approach is used for training. We then further split the training set to estimate model hyperparameters $\lambda$ and $\sigma$ using five-fold cross-validation.
We performed an extensive 2-step grid-search to find the hyperparameters which minimize the $R^2$ score for each separate location. In the first step a coarse-grained grid was used to identify the area in which hyperparameters were acceptable, in the second step we zoomed in around such area to further refine the hyperparameter values.



\bibliographystyle{elsarticle-harv} 
\bibliography{biblio.bib}

\end{document}